# The Business of Selling Electronic Documents

Manuel Oriol
University of York
manuel@cs.york.ac.uk

**Abstract**
The music industry has huge troubles adapting to the new technologies. As many pointed out, when copying music is essentially free and socially accepted it becomes increasingly tempting for users to infringe copyrights and copy music from one person to another. The answer of the music industry is to outlaw a majority of citizens. This article describes how the music industry should reinvent itself and adapt to a world where the network is ubiquitous and exchanging information is essentially free. It relies on adapting prices to the demand and lower costs of electronic documents in a dramatic way.

## 1. Introduction

The mutation of the traditional distribution channels has an impact on all sectors of the industry. Because of the diversity of available products it is now faster, and sometimes cheaper, to order objects over the Internet and have them delivered directly at home than to go to a shop that would order them. The same fact is even more prevalent with digital goods. Buying a program online, buying music online, buying movies online is much faster than going to a shop and come back. Buying – or renting – digital goods is however usually not cheaper than buying them in stores. Because a digital good does not need the same packaging this single fact is unacceptable to consumers.

With similar ease it is possible to obtain the same goods for free by breaking a law that is broken on a daily basis. Moreover, by downloading some pirated digital good over the Internet users get rid of some of the limitations put in place to guarantee a revenue stream to major players in the recording industry.

Some people are however inclined to pay for digital goods. Early adopters are a good example of that. For example most video games are making most of their profit in the first few weeks following their publication. Other examples of customers willing to pay are people who strongly believe in being honest. The latter type is however increasingly tempted by unlawful ways of acquiring digital content.

The present approach builds on the common belief that what is rare is expensive and what is widespread is cheap. The basic idea is that if a digital good is already owned by many, then acquiring it should be very cheap, while the first people to acquire it might pay a higher amount of money to use them. Once the digital good is so widespread that millions of people bought it, it does not make sense to pay for it anymore and the digital good should become free as in effect it belongs now to the global culture.

The main advantage of such a model of distribution is that it maps quite well current practices and reduces dramatically prices for digital goods, as there is no production of physical objects involved. It also has the advantage of integrating emergent countries where people cannot afford to pay high price tags but can afford to wait for a lower price. Potentially such a system would reduce the margins of the black market resellers and make such businesses not ludicrous enough to compensate the inherent risks.

Section 2 presents a criticism of the existing structures, Section 3 presents the distribution model for digital goods, Section 4 presents balancing mechanism for the system to finance creation and foster new talents into the digital culture. Conclusions are eventually presented in Section 5.

## 2. Criticism of the existing structures

Currently, companies play the role of editors, they select artists and content providers and pay them to use their production. When the costs are expensive they even finance what they think might sell well and provide the means to help the production like for movies and music. The super-productions are now incredibly powerful and produce results for the masses. No independent producer can really compete with such powers.

It is not possible to copy digital goods and make profit out of it due to laws that were passed to protect the author and the production companies. The license of use granted to the buyer is limited because of the fear of the big companies that people might not pay to get the digital goods. With the arrival of the numerical world the objects sold are becoming easily reproducible. To counter that and help the music industry maintaining their margins, some countries put taxes on storage media, making people who do not infringe the law pay twice: when they buy the storage media and when they buy de record. Each time the system is put in question, the companies show well-known, wealthy artists that defend this process that brought them to the top.

In parallel, computers become cheaper and cheaper and the combination of software and hardware needed to produce professional-grade artistic material has become so cheap that the artists are supposed to own at least stripped down versions of them. The cost of production has been relocated to the artist himself almost completely.

The market of third-world countries is overflown by illegal records of movies and music. In the same way, peer-to-peer technologies and software specialized in removing protections put in place by the media companies have seen a huge increase in use.

Some countries realize that if everybody who once committed the felony of copying illegally information and art pieces would be put to jail, they would have to put more than half of their population there. These countries then try to pass laws that are fighting back such as the law on paying a flat fee to download anything in France. These laws, like in France, are then actively fought by the media companies that eventually win their fights.

Why are so many people willing to copy information without paying the fees? The business model itself does not correspond to what people expect fair use should be. People think that after enough people have bought the information, it is only fair that it becomes de facto a freely available product. What is still preventing people to copy all the music they want? Morality only.

How does the economy of a CD works now? What do the artists earn? Apparently, on a $15 CD, $5 go to the store, $3 to the distributor and the rest is shared between the label and the artist. Because the music label advances the fees to effectively produce the CD, it reimburse these fees first and then pays the musicians.[1] In the end, musicians receive an average share of 12-13% of the 7 remaining dollars. As an example, a band recording an album that sells a million of exemplars will receive $910'000. This means that the information producer, the real artist receives less than 6% of the actual price of a CD! Comparatively, when they go on tour and make live performances the artists often make as much or more money and this can become a significant part of their revenue.

People do not accept this model where the information producers makes such a small part of the profits. With this model, asking for a morally acceptable behavior from the consumer while the producers seem to have morally questionable motivations.

## 3. Proposed model

What people can do with Information is changing. The main change is that Internet has now a bandwidth big enough to support any important data transfer in matters of minutes rather than days. This is at the heart of a revolution that the media companies are trying to fight by defending their way of producing and selling media. This is a fundamental mistake that shows how little the decision-takers understand the new media.

It is needed to reward better the information producers and after some benefits the media should become public domain. It should also be possible for people to find some morality back by paying for the information. The proposition made here is the model that would best answer the dilemma at hand.

---

[1] Found on http://weeklywire.com/ww/08-24-98/memphis_cvr.html, November 6th, 2007

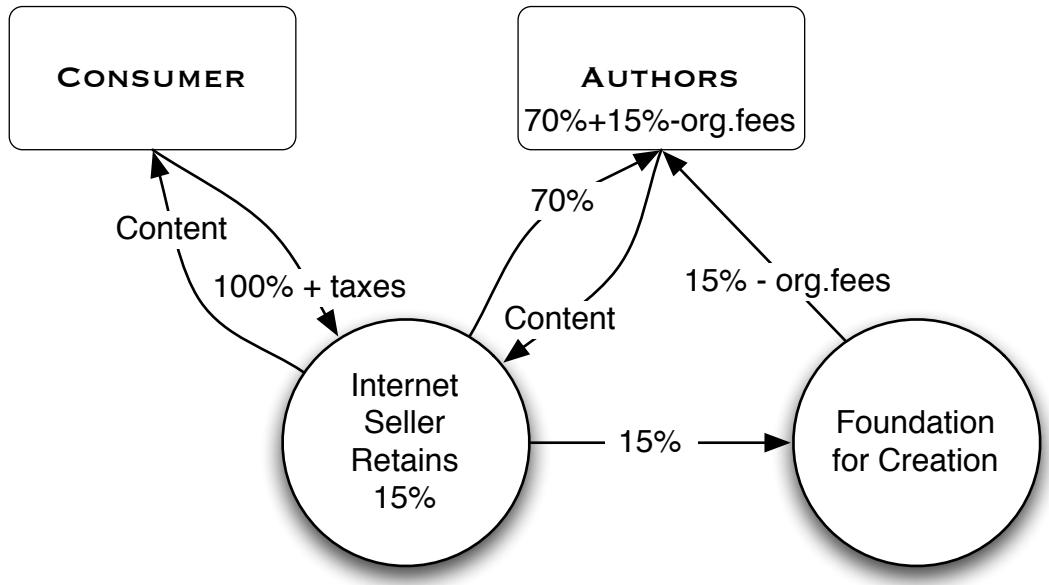

**Fig. 1: Repartitions of money and content**

The main idea is that when more people have bought a piece of information, its value decreases. It decreases down to a point where it is free. At that point it becomes freely available on the peer-to-peer networks and anybody can have it, it becomes then a part of the human culture. People who bought the piece of information in the first phase can find it rewarding that they contributed to the increase of the human culture. The process should also be completely transparent with a clear policy. At the core of the system, the artist sets up the amount of money that it would like to receive for a given information. It could be for free or it could cost a fixed amount of money. People should have the right to give the digital good to other people that they can touch physically, so that friends and family could exchange information when meeting. At any point the buyers should know at which end of the ladder they are and how many more buyers are needed to make it free of charge.

The service provider should receive a fixed percentage of the sales (see Fig. 1) and to encourage young artists a nonprofit organization should also be used to allocate grants that would let young artists produce their own music, movies, books etc. As a first approximation it seems reasonable that the company in charge of distributing and advertising artists would receive 15% of the sale and that the foundation would receive 15% as well. Information producers should then receive 70% of the price of the information. This is in line with what Apple proposes through the iTunes AppStore.

To verify the validity of the model it is needed to test it on a concrete example. To make it work, it is important to understand that to succeed, this system has to reduce drastically the costs of producing music. A factor of 10 seems to be the kind of reduction that would draw sufficient interest from consumers. Let now consider that an album would cost $1.5 at the beginning of its sales. Let say that the formula applied to find the price paid by a buyer n would be of the form:

$$price(n) = \text{roundup}(\gamma e^{-\beta n} - \alpha)$$

Where $\alpha$ is the floor price below which it is considered to be free (here we consider that $\alpha=.01$ for representing one cent), $\gamma-\alpha$ represents the starting price of the album (in our case $\gamma=1.51$), $\beta$ represent a way to act on the benefits made by the artist. An approximation of the money earned is the integral between 0 and when the price is null. this happens when price(n) = 0. This is the case for n= $(\ln(\alpha/\gamma))/(-\beta) \simeq 5/\beta$. The earnings can then be approximated to:

earnings ($\beta$) = PRICE(5/$\beta$) - PRICE(0) where PRICE(n)= - $\gamma/\beta$ * $e^{-\beta n}$ - $\alpha n$
earnings ($\beta$) = - 1.51/$\beta$ * $e^{-5}$ - .05 $\beta$ + 1.51/$\beta$ $\simeq$ 1.51/$\beta$ (if $\beta$ is very close to 0)

In the case of a platinum record (1'000'000 albums sold) then $\beta$ = 1/200'000 and earnings are close to $300'000. The inverse calculation is also possible and for an album to give $910'000 to the band should have earnings of $1'300'000 which means that $\beta$=1/860'000 and the number of people that buy it should be 4'300'000. Most of those people would nevertheless buy the album for less than one dollar and one could expect that these prices would drive more attention to the album itself. While these values are a very rough approximation, they look reasonable enough. Artists should be offered the possibility to choose the starting price and how much money they want to make out of the generated information.

## 4. Informal evaluation

The current model of the majors works because it favors the emergence of new artists by funding them. The new business model proposed here does the same thing but in a different way: it uses a non-profit organization to take up that role. The 15% of the earnings that go to the foundation should enable artists to record songs and albums. These should be allocated in priority to the artists that made the record that generated the revenue stream to create their next album. The rest of the money should be redistributed to unknown artists that need help to create their own albums if they did not have any previous revenues. Due to current costs – $50'000 for a recognized band, $10'000 for newcomers – a band that generates $1'000'000 would be able to record another album and 10 lower profile artists would be helped to create their own.

Because the model is applied only by artists willing to do so, it is likely that major music businesses would still survive for a long time alongside with artists using the new system. Shops selling discs would not disappear, they would change the way they do business. It is likely that they would have automata that would let them burn CDs in a convenient manner and with a better quality than people would do it at home. Shops would also probably have to focus on special events that would drive people to shops rather than only showing discs and taking money from customers.

## 5. Related Work

Nobody so far proposed a similar model. The only model that big publishers proposed is progressive: the price is cheap at first and then goes up. This is obviously unappealing to people as piracy will become more attractive in time. The other solution offered by authors is simply to destroy copyright [1] and to rely on live performances to generate revenues. This is clearly not adapted to a global economy.

## 6. Conclusions

The model outlined here is only a first step and should be refined and modified to adapt to particulars. Several approximations have been made to make the model simple. Taxes are not accounted but they should probably be paid at each sale. Overhead is grossly evaluated and the exact percentages received by each organization should be reevaluated to fit exact needs. A lot of details are not taken into account like the decision process of allocating money by the foundation, the exact decreasing function for the price could be something else, etc.

This model, however, is more likely to succeed than any other because it maps the way people use and share the electronic media. It has the advantage to make legal a de facto standard where the electronic version of a media is actually free=    . It is also very egalitarian as poorer people can wait for a given good to be less expensive and then buy it. To conclude, the model also makes each buyer a better citizen of the world because it participates to the general good by bringing each medium she buys a little closer to belong to the world community.

In our opinion this is a good business model because it aims at making reasonable amounts of money from established practices and transforms the perception of buyers from being stupid to being good citizens and the perception of pirates as being impatient and malevolent.